\documentstyle[twoside, epsfig]{article}

\input ibvs2.sty

\begin{document}
\IBVShead{5xxx}{19 Jan 2015}

\IBVStitle{AR Ser: photometric observations of a Blazhko star}

\IBVSauth{Michel Bonnardeau$^1$; Franz-Josef (Josch) Hambsch (HMB)$^2$}

\IBVSinst{MBCAA Observatory, Le Pavillon, 38930 Lalley, France, email: arzelier1@free.fr}
\IBVSinst{ROAD Observatory, 12 Oude Bleken, Mol, 2400, Belgium}

\SIMBADobjAlias{AR Ser}
\IBVStyp{GCVS}
\IBVSkey{photometry}
\IBVSabs{Photometric observations in 2010-2014 of the RR Lyrae star AR Serpentis are presented and analyzed. Two Blazhko modulations of comparable amplitude are detected, with the periods 89 and 108 days, and with evidence for irregularities.}

\begintext

We observed AR Serpentis (RA=15h33min30.8s DEC=+2\deg46\arcm38\arcs (2000.0) average V mag=12.0) between 2010 and 2014, with Johnson V filters and CCD cameras. 2195 photometry measurements over 58 nights were gathered in 2010-2014 with a 20 cm telescope located in France (MB), and 4599 measurements over 74 nights in 2014 with a 40 cm telescope in Chile (HMB). 

For the differential photometry, the comparison star is UCAC4 464-053185 with a V magnitude of 11.551. Because the instruments are different, there is a magnitude offset between the two observers. Owing to 3 pairs of overlapping times-series with the two setups (as shown in Figure 1), this offset is evaluated to be 20 mmag, that is added to the measurements obtained with the 20 cm telescope. 
 
\IBVSfig{6cm}{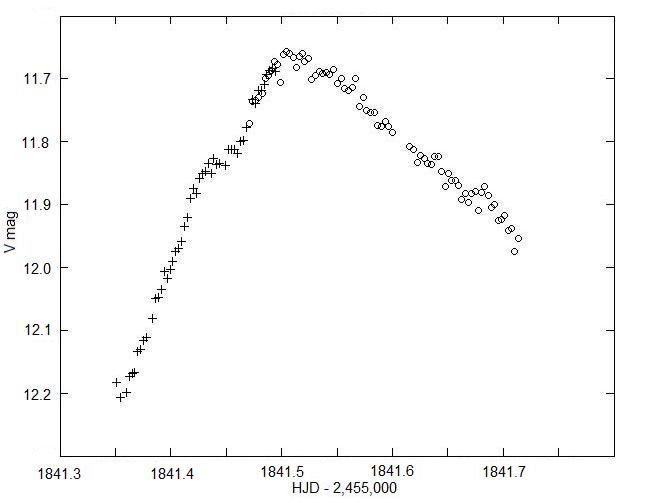}{An example of a pair of overlapping times-series. Cross: data with the 20 cm telescope, circle: with the 40 cm one.}

The data are analyzed with the PERIOD04 software program (Lenz \& Breger, 2005) which provides simultaneously sine-wave fitting and least-squares fitting algorithms. This yields the pulsation frequency: \\$F_{p} = 1.7385671 \pm 3.9.10^{-6} \; day^{-1}$ 
\\(or the pulsation period: $P_{p} = 0.5751863 \; day \pm 0.12 \; s$).

There is no evidence for a variation of this period during our observations. The ephemeris for the maxima of the pulsation is then:\\$t(n) = 2,456,135.181 + nP_{p} \; HJD$

Owing to the PERIOD04 software program, the data are fitted with a sine-wave function of  time t, with the number of harmonics of up to the 7th order: 
\\
$ f(t) = z +\Sigma_{i=1}^{8}A_{i}\sin [2\pi(F_{i} t+\Phi_{i})] $
\\with $ z = 11.9597 \pm 0.0014 \; mag$ and the other parameters given in Table 1, components F1-F8. 

The resulting phase plot is shown in Figure 2 and the residuals of the observations from the f(t) function in Figure 3.

\IBVS2fig{7.8cm}{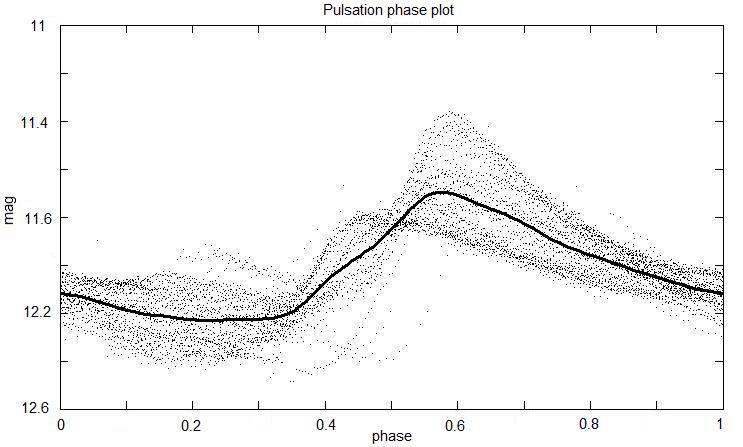}{Dots: the observations, Solid line: the f(t) function. The phase origin is arbitrary. }{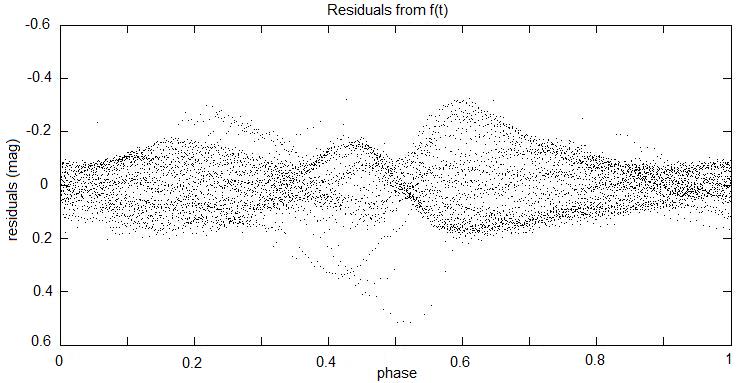}{The difference between the observations and the f(t) function.}

The deviations around the fit f(t) are due to the Blazhko effect.

The Blazhko modulation is expected to show up as side peaks around multiples of the pulsation frequency $F_{p}$ in the Fourier spectrum (Breger \& Kolenberg, 2006 and Szeidl \& Jurcsik, 2009). The two strongest signals in the residuals of the pulsation (prewhitening) correspond to the frequencies $F_{B1} = 1/89 \; day^{-1}$ and $F_{B2} = 1/109 \; day^{-1}$. Peaks are clearly seen at $nF_{p} \pm F_{B1}$ and $nF_{p} \pm F_{B2}$ with n=1, 2, 3, as shown in Figures 4, 5, 6. Hence, AR Ser has two Blazhko modulations.

There are many fainter peaks in the spectra and we refrain to interpret them. However there is a possibility for signals at $nF_{p} \pm 2F_{B2}$ (see Figures 4, 5, 6), although fitting the data including them does not much improve the residuals. Such quintuplets may imply a magnetic origin for the modulation (Shibahashi, 2000) or may be due to a non-radial pulsation (Dziembowki \& Mizerski, 2004). The first star where a Blazhko effect was discovered with quintuplets was RV UMa (Hurta et al., 2008) and a number of them have been found since, especially owing to satellite observations (Chadid et al., 2010).

The observations are not evenly distributed, with most of them concentrated over a time interval of about $1/F_{B1}$ (the data obtained from Chile). We checked that the signals at $F_{B1}$ are not spurious by computing the Fourier spectrum of the data obtained from France only, that span 5 years: the strongest signals are still at $F_{B1}$.

Using the PERIOD04 software program, the observations are then fitted with the frequencies $nF_{p} \pm F_{B1}$ and $nF_{p} \pm F_{B2}$. The results are shown in Table 1, components F9 through F20.

The data from Chile have a very dense coverage. Their Fourier spectrum was then searched for a high frequency Blazhko modulation, with a negative result.

The residuals of the observations and of this modeling (the F(t) function, see below) were searched for signals at $F_{B1}$ and $F_{B2}$, in the low frequency end of their Fourier spectrum, with negative results. The spectrum was also searched for half-integer multiples of the pulsation frequency $F_{p}$ (this is connected to the period doubling, see Szab\'{o}, 2014),  with negative results.

The fit function F(t) is:
\\
$ F(t) = Z +\Sigma_{i=1}^{20}A_{i}\sin [2\pi(F_{i} t+\Phi_{i})] $
\\with $ Z = 11.95590 \pm 0.00057 \; mag$.

\IBVSfig{12cm}{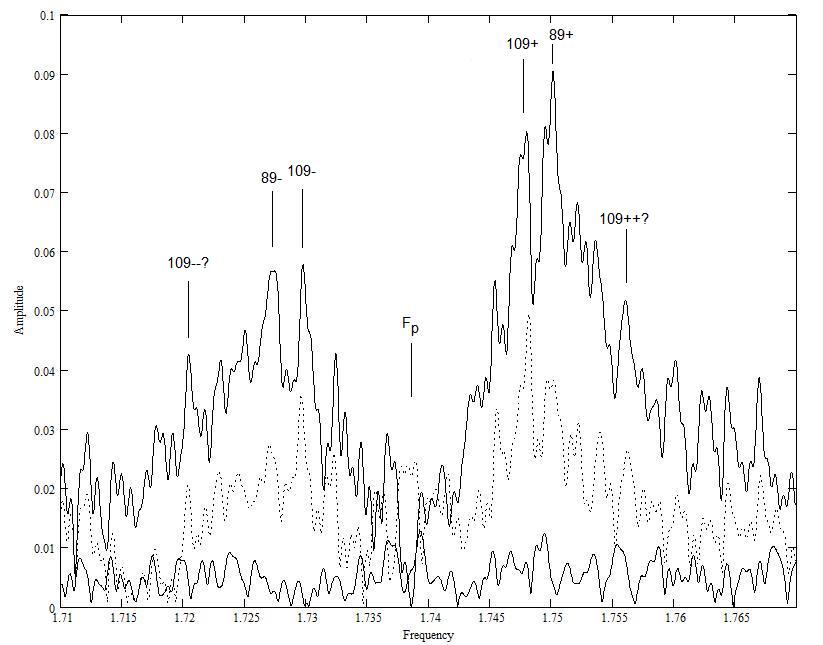}{The Fourier spectra of the residuals around $F_{p}$. The upper solid line is after prewhitening with the $F_{p}$ pulsation. The triplet at $F_{B1}$ is noted 89+ and 89-, the one at $F_{B2}$, 109+ and 109- (there is a hint for a quintuplet with $F_{B2}$, hence the 109++ and 109- -). The middle dotted line is after prewhitening with both the $F_{p}$ pulsation and the $F_{B1}$ modulation: the triplets at $F_{B2}$ are clearly visible. The bottom solid line is after prewhitening with the pulsation and the modulations at $F_{B1}$ and $F_{B2}$: only noise is left.} 

\IBVS2fig{7cm}{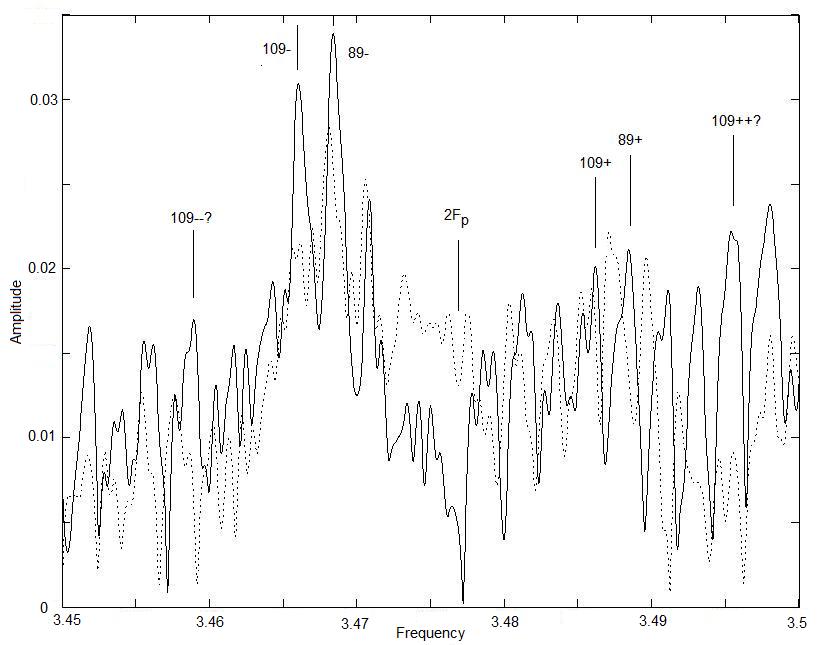}{The Fourier spectrum around $2F_{p}$. Solid line: prewhitening with the pulsation only, Dotted line: with the pulsation and the $F_{B1}$ modulation.}{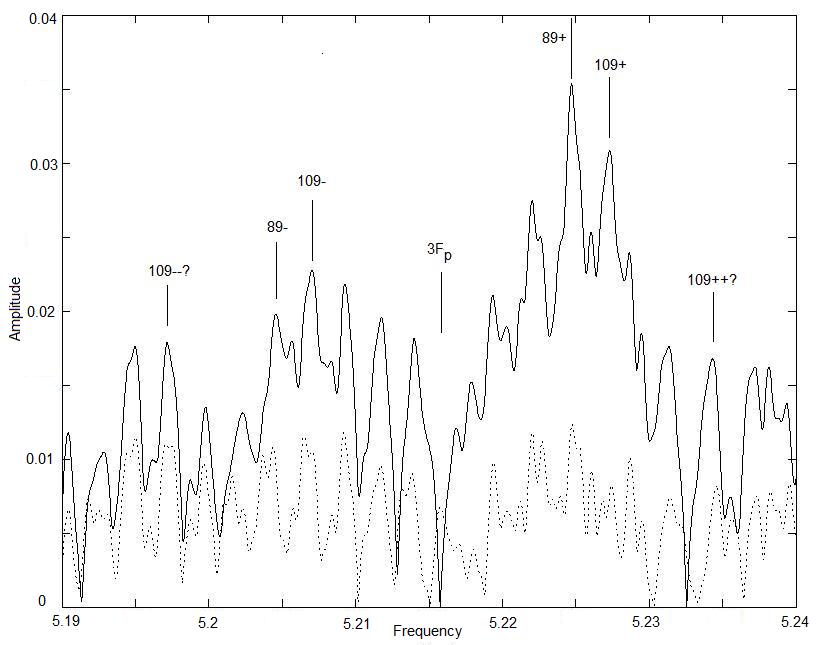}{The same as Fig 5 for $3F_{p}$.}

The difference between the F(t) and the f(t) represents the Blazhko modulations. It is shown in Figure 7.

The two Blazhko periods may be computed from the F9, F10, F13, F14, F17 and F18 components of Table 1 for the first period, and from the other components starting from F11 for the second period. The average and standard deviations are:
\\
$1/F_{B1} = 89.1 \pm 1.3 \; days$
\\
$1/F_{B2} = 109.6 \pm 2.6 \; days$.

\vskip 0.1cm
\centerline{Table 1: Sinusoidal decomposition with PERIOD04.}
\begin{center}
\begin{tabular}{|l|l|l|l|l|l|l|l|}
\hline   Name & Component  & Frequency  & Uncertainty  & Amplitude  & Uncert.  & Phase  & Uncert.    \\
  &  & $F$ & on $F$ & $A$ & on  $A$ & $\Phi$ & on $\Phi$   \\
\hline  F1 & $F_{p}$ & 1.7385671 & 3.9e-006 & 0.2155 &	0.0018 &	0.1017 &	0.0013   \\
 
\hline  F2 &   $2F_{p}$ & & & 0.0726 &	0.0018 &	0.6056 &	0.0042 \\ 
\hline  F3 &  $3F_{p}$ & &  & 0.0206 &	0.0019 &	0.078 &	0.013  \\ 
\hline  F4 & $4F_{p}$ & &   & 0.0063 &	0.0019 &	0.525 &	0.061  \\ 
\hline  F5 &  $5F_{p}$ & &  & 0.0052 &	0.0018 &	0.730 &	0.043  \\ 
\hline  F6 &  $6F_{p}$ & &  & 0.0069 &	0.0016 &	0.356 &	0.039  \\ 
\hline  F7 & $7F_{p}$ & &   & 0.0066 &	0.0021 &	0.833 &	0.046  \\ 
\hline  F8 &  $8F_{p}$ & &  & 0.0049 &	0.0020 &	0.380 &	0.053   \\ 
\hline  F9 & $F_{p} + F_{B1}$ & 1.7500584 &	7.8e-006 &	0.0762 &	0.0018 &	0.5796 &	0.0039 \\ 
\hline  F10 & $F_{p} - F_{B1}$ & 1.727451 &	1.9e-005 &	0.0295 &	0.0017 &	0.8223 &	0.0092 \\ 
\hline  F11 & $F_{p} + F_{B2}$ & 1.7481126 &	9.9e-006 &	0.0557 &	0.0019 &	0.4206 &	0.0050 \\ 
\hline  F12 & $F_{p} - F_{B2}$ & 1.7296166 &	9.5e-006 &	0.0413 &	0.0018 &	0.6841 &	0.0048 \\ 
\hline  F13 & $2F_{p} + F_{B1}$ & 3.488479 &	1.6e-005 &	0.0276 &	0.0013 &	0.2376 &	0.0069 \\ 
\hline  F14 & $2F_{p} - F_{B1}$ & 3.466037 &	1.5e-005 &	0.0299 &	0.0014 &	0.2942 &	0.0069  \\ 
\hline  F15 & $2F_{p} + F_{B2}$ & 3.485996 &	3.4e-005 &	0.0168 &	0.0013 &	0.454 &	0.016  \\ 
\hline  F16 & $2F_{p} - F_{B2}$ & 3.468079 &	1.6e-005 &	0.0242 &	0.0013 &	0.241 &	0.047  \\ 
\hline  F17 & $3F_{p} + F_{B1}$ & 5.226968 &	2.1e-005 &	0.0240 &	0.0013 &	0.787 &	0.011 \\ 
\hline  F18 & $3F_{p} - F_{B1}$ & 5.204693 &	2.5e-005 &	0.0128 &	0.0013 &	0.643 &	0.012   \\ 
\hline  F19 & $3F_{p} + F_{B2}$ & 5.224864 &	2.5e-005 &	0.0121 &	0.0013 &	0.830 &	0.013  \\ 
\hline  F20 & $3F_{p} - F_{B2}$ & 5.206496 &	2.6e-005 &	0.0143 &	0.0012 &	0.077 &	0.012    \\ 
\hline 
\end{tabular}
\end{center}
%\vskip 0.3cm

AR Ser was reported in the literature having one Blazhko modulation close to $F_{B2}$ (Firmanyuk, 1977, Kolenberg et al., 2008) and also an uncertainty pulsation of 63 days  (Wils et al., 2006), not seen in our data. We observe it with two modulations of comparable amplitude. The first Blazhko star discovered as having two modulations of comparable amplitude is CZ Lac (S\'{o}dor et al., 2011). Such stars are not very common in ground-based observations (Skarko, 2014) although they seem to be ubiquitous in satellite observations (Benk\H{o} et al., 2014).

\IBVSfig{12cm}{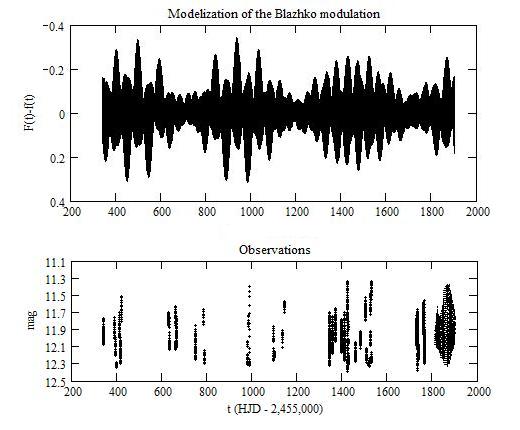}{ The Blazhko modulations and the observations.}
\IBVSfig{7cm}{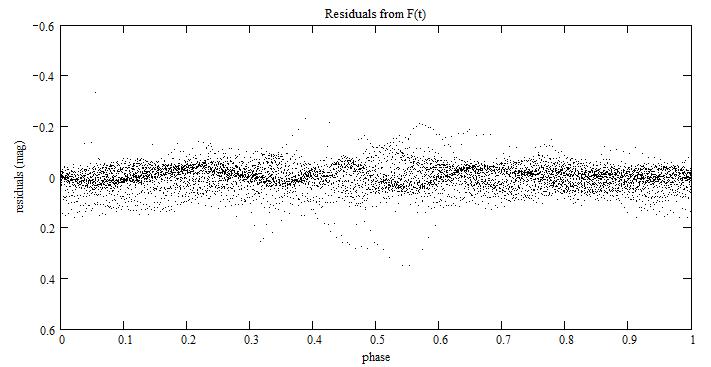}{ Residuals of the observations and of the F(t) function.}

The ratio of the two frequencies may be expressed as the ratio of two small integers, $F_{B1}/F_{B2}=6/5$, a common occurrence for Blazhko stars with two modulations (Skarka, 2014, Benk\H{o} et al., 2014, S\'{o}dor et al., 2011).

The residuals of the observations and of the F(t) function are shown in Figure 8. Although they are much improved compared to Figure 3, there are a few time-series that do not fit the model F(t) and are out of phase or with too large amplitudes. Such discrepant observations appear suddenly, that means the time series obtained a few weeks or days before or after fit the model. This suggests irregularities or glitches, which is a behavior observed in many Blazhko stars (Szab\'{o}, 2014).

% \vskip 3cm

\references

Benk\H{o} J.M., Plachy E., Szab\'{o} R., Moln\'{a}r L. and Koll\'{a}th Z., 2014, \textit{ApJS} \textbf{213} 31.

Breger M. and Kolenberg K., 2006, \textit{A\&A} \textbf{460} 167.

Chadid M., Benk\H{o} J.M., Szab\'{o} R. et al., 2010, \textit{A\&A}, \textbf{510} A39.  

Dziembowki W.A. and Mizerski T., 2004, \textit{Acta Astronomica} \textbf{54} 363.
 
Firmanyuk B.N., 1977, \textit{IBVS} 1245.

Hurta Zs., Jurcsik J., Szeidl B. and  S\'{o}dor \'{A}., 2008, \textit{A.J.} \textbf{135} 957.

Kolenberg K., Guggenberger E. and Medupe T., 2008, \textit{Comm. Asteroseismology} \textbf{153} 67.

Lenz P. and Breger M., 2005, \textit{Comm. Asteroseismology} \textbf{146} 53.

Shibahashi H., 2000 in \textit{The Impact of Large-Scale Surveys on Pulsating Star Research}, \textit{ASP Conf. Series} \textbf{203} 299, Szabados L. and Kurtz D.W., eds.

Skarka M., 2014, \textit{A\&A} \textbf{562} 90.

S\'{o}dor \'{A}., Jurcsik J., Szeidl B., V\'{a}radi M., Henden A., Vida K., Hurta Zs., Posztob\'{a}nyi K., D\'{e}k\'{a}ny I. and Szing A., 2011, \textit{MNRAS} \textbf{411} 1585.

Szab\'{o} R., 2014, in \textit{Precision Asteroseismology, Celebrating the Scientific Opus of Wojtek Dziembowski}, \textit{Proc. IAU Symposium} \textbf{301} 241, Chaplin W., Guzik J.A., Handler G. and Pigulski A., eds.

Szeidl B. and Jurcsik J., 2009, \textit{Comm. Asteroseismology} \textbf{160} 17.

Wils P., Lloyd C. and Bernhard K., 2006, \textit{MNRAS} \textbf{368} 175.

\endreferences

\end{document}